\documentclass[usenatbib]{mn2e}
\usepackage{graphicx,amsmath,multirow,amssymb}
\usepackage{subfig}
\usepackage{natbib}
\newcommand{\comment}[1]{}

\def\simgt{\lower.5ex\hbox{$\; \buildrel > \over \sim \;$}}
\def\simlt{\lower.5ex\hbox{$\; \buildrel < \over \sim \;$}}

\newcommand{\Msun}{\ensuremath{{\rm M}_{\sun}}}

\title[Galactic planetary nebulae ]{The evolution of Galactic planetary nebula progenitors 
through the comparison of their nebular abundances with AGB yields.}
\author[Ventura et al.]{P. Ventura$^1$, L. Stanghellini$^{2}$, F. Dell'Agli$^{3,4}$,
D. A. Garc\'{\i}a--Hern\'andez$^{3,4}$, 
 \\
$^1$INAF -- Osservatorio Astronomico di Roma, Via Frascati 33, 00040, Monte Porzio Catone (RM), Italy \\
$^2$National Optical Astronomy Observatory, 950 N. Cherry Avenue, Tucson (AZ) 85719, USA \\
$^{3}$Instituto de Astrof\'{\i}sica de Canarias, E-38205 La Laguna, Tenerife, Spain \\
$^{4}$Departamento de Astrof\'{\i}sica, Universidad de La Laguna (ULL), E-38206 La Laguna, Tenerife, Spain\\
}

\begin{document}

\date{Accepted, Received; in original form }

\pagerange{\pageref{firstpage}--\pageref{lastpage}} \pubyear{2017}

\maketitle

\label{firstpage}

\begin{abstract}
We study the chemical abundances of a wide sample of 142 Galactic planetary nebulae (PNe)
with good quality observations, for which the abundances have been derived more or less 
homogeneously, thus allowing a reasonable comparison with stellar models. 
The goal is the
determination of mass, chemical composition and formation epoch of their progenitors, through 
comparison of the data with results from AGB evolution. The dust properties of PNe, when 
available, were also used to further support our interpretation. 

We find that the majority ($\sim60\%$) of the Galactic PNe 
studied has nearly solar chemical composition, while $\sim40\%$ of the sources 
investigated have sub-solar metallicities. 
About half of the PNe have carbon star progenitors, in the $1.5~M_{\odot} < M < 3~M_{\odot}$  
mass range, which have formed between 300 Myr and 2 Gyr ago. The remaining PNe are
almost equally distributed among PNe enriched in nitrogen, which we interpret as the progeny 
of $M > 3.5~M_{\odot}$ stars, younger than 250 Myr, and a group of oxygen-rich PNe, 
descending from old ($> 2$ Gyr) low-mass ($M < 1.5~M_{\odot}$) stars that never became 
C-stars.

This analysis confirms the existence of an upper limit to the amount of carbon
which can be accumulated at the surface of carbon stars, probably due to 
the acceleration of mass loss in the late AGB phases. The chemical composition of the
present sample suggests that in massive AGB stars of solar (or slightly sub-solar)
metallicity, the effects of third dredge up combine with hot bottom burning, resulting
in nitrogen-rich - but not severely carbon depleted - gaseous material to be ejected.
\end{abstract}

140.252.118.146
\begin{keywords}
Planetary Nebulae: individual -- Stars: abundances -- Stars: AGB and post-AGB. Stars: carbon 
\end{keywords}

\section{Introduction}
The stars of mass below $\sim 8\Msun$, after the end of core helium burning, enter the
AGB phase, during which a CNO burning shell provides the thermonuclear energy supply 
almost entirely. Periodically, a helium-rich layer above the degenerate core is ignited
in conditions of thermal instability, hence the name "thermal pulse" (hereinafter TP), 
used to define these episodes.

Despite relatively short in comparison with the core hydrogen and helium burning
phases, the AGB evolution proves extremely important to understand the feedback of these
stars on their host system, because it is during this phase that most of the mass 
loss occurs, with the consequent pollution of gas and dust of the interstellar medium.

A full comprehension of the AGB evolution proves crucial for a number of astrophysical
contexts, such as the determination of the masses of galaxies at high redshifts \citep{maraston06}, 
the formation and chemical evolution of galaxies \citep{romano10, santini14}, the dust 
content of high-redshift quasars \citep{valiante11}, and the formation of multiple 
populations of stars in globular clusters \citep{ventura01}.

The modelling of the AGB phase has made significant progresses in the recent years, with 
the description of the dust formation process in the winds, coupled to the 
evolution of the central star \citep{fg06, nanni13, nanni14, ventura12a, 
ventura12b, marcella13, ventura14a}.

The results are still rather uncertain though, because of the poor knowledge of convection
and mass loss, two physical mechanisms having a strong impact on the physical and
chemical features of the AGB evolution \citep{karakas14, vd05}. Therefore, the comparison 
with the observations is extremely important to allow a qualitative
step towards an exhaustive knowledge of the main properties of these stars.

The Magellanic Clouds (MC) have been so far the most investigated environments to this aim, 
owing to their relatively short distances (51 kpc and 61 kpc respectively, for the Large 
and Small Magellanic Cloud, Cioni et al. 2000; Keller \& Wood 2006) and the low reddening 
($E_{B-V}=0.15$~mag and 0.04 mag, respectively, for the LMC and SMC, Westerlund 1997).
The near- and mid-infrared observations of stars in the MC, particularly suitable to
study AGB stars, given the cool (below $\sim 4000$K) surface temperatures of these objects,
have been extensively used to constrain the AGB models \citep{izzard04, girardi07, 
martin07, riebel10, riebel12, srinivasan09, srinivasan11, boyer11, boyer12}. These studies 
have been completed by more recent investigations, focused on dust production expected 
from this class of stars, and the relative effects on the infrared colours,  
\citep{flavia14, flavia15a, flavia15b, ventura15, ventura16, nanni16}.

A complementary approach to infer valuable information on the evolution
of AGB stars is offered by the study of PNe. The chemical composition of these objects
reflects the final surface chemistry, at the end of the AGB phase, and is determined by the
combination of the two mechanisms potentially able to change their surface chemistry,
namely third dredge up (TDU) and hot bottom burning (HBB). The determination of the
abundances of the individual species in the PN winds provides a unique tool to understand
the efficiency of the two mechanisms. 

Motivated by these arguments, we have recently started a research project aimed at interpreting
the observed sample of PNe based on recent AGB models, accounting for the formation of
dust in the circumstellar envelope. In the first two papers of this series we focused
on the PNe population of the LMC (Ventura et al. 2015b, hereinafter paper I) and of the 
SMC (Ventura et al. 2016b, paper II). These works allowed the characterization of the
observed PNe in terms of the mass and metallicity of the progenitors.

Here we extend this study to the population of Galactic PNe, whose $\alpha$-element 
abundances indicate a wider range of progenitor metallicities. Our aim is
twofold: a) we attempt to characterize the individual PNe observed and to identify the
progenitors; b) we test AGB evolution and the dust formation process against a different 
and more complex environment than that of Magellanic Cloud PNe.

The paper is organised as follows: section 2 provides a description of the most
important physical and chemical input used to model the AGB phase; in section 3 we discuss
the modification of the surface chemical composition of AGB stars, and the expected,
final abundances of the various chemical species; the interpretation of the two samples
of Galactic PNe studied in the present work is addressed in section 4; in section 5 we
discuss the results obtained on the basis of the dust features detected in the spectra
of some of the PNe observed;the conclusions are given in section 6.  

\section{AGB modelling}
Our aim is to interpret the abundances of specific Galactic PN samples on the basis of AGB 
models of different mass and chemical
composition. Our goal is to deduce the mass and the metallicity of the progenitor of the
individual PNe observed, by comparing the abundances of the various chemical species
at the end of the AGB phase with the values derived from the observations. Before
entering this detailed comparison, we provide a brief description of the AGB models
adopted.

The evolutionary sequences have been calculated by the ATON code for stellar evolution.
The details of the numerical structure of the code are extensively discussed in
\citet{ventura98}, whereas the most recent updates are presented in \citet{ventura09}.
Here we provide a short description of the physical and chemical input most relevant
to this work.

\subsection{Convection}
The temperature gradient within regions unstable to convective motions is found by
the full spectrum of turbulence (FST) model for convection \citep{cm91}. Mixing of
chemicals and nuclear burning are treated simultaneously, by means of a diffusive-like
approach. During the two major core burning phases we assume that convective velocities
decay exponentially from the border of the core within radiatively stable regions,
with an e-folding distance of $0.02H_p$; this choice is motivated by a calibration
of core overshooting based on the observed extension of the main sequences of open
clusters, given in \citet{ventura98}. During the AGB phase we assume overshoot from
the base of the envelope and from the borders of the convective shell which forms
at the ignition of each TP; in this case the e-folding distance of convective velocities
is $0.002H_p$, according to a calibration based on the luminosity function of carbon
stars in the LMC, given in \citet{ventura14a}.

\subsection{Mass loss}
We adopt the formalism by \citet{blocker95} to describe mass loss of oxygen-rich AGB
stars. Blocker's formula consists in the canonical Reimers' mass loss rate 
multiplied by a power of the luminosity, $L^{2.7}$. The free parameter entering the
Reimers' rate was set to $\eta_R=0.02$, following the study on the luminosity function
of lithium-rich stars in the MC, by \citet{ventura00}. 

For what regards carbon stars, we adopt the results from hydrodynamical models of
carbon stars, published by \citet{wachter02, wachter08}.

\subsection{Opacities}
Radiative opacities are calculated according to the OPAL release, in the version 
documented by \citet{iglesias}. The molecular opacities in the low-temperature 
regime ($T < 10^4$ K) are calculated by means of the AESOPUS tool \citep{marigo09}. 
The opacities are constructed to follow the changes of the envelope chemical composition, 
in particular carbon, nitrogen and oxygen individual abundances.

\subsection{Chemical composition}
The AGB models used here have metallicities $Z=10^{-3}, 2\times 10^{-3}, 4\times 10^{-3}, 
8\times 10^{-3}$, $Z=0.014$, $Z=0.018$, $Z=0.04$. In the $Z=1,2 \times 10^{-3}$ models
we assume the mixture by \citet{gs98}, with an $alpha-$ enhancement $[\alpha/Fe]=+0.4$;
the $Z=4,8 \times 10^{-3}$ models were calculated with the \citet{gs98} mixture and an
$\alpha-$ enhancement $[\alpha/Fe]=+0.2$; the $Z=0.014$ models are based on the 
solar-scaled mixture by \citet{lodders03}; finally, the $Z=0.018$ and $Z=0.04$ models have a 
solar-scaled mixture, with the distribution by \citet{gs98}. The initial abundances
of the chemical species mostly used in the present work for the various metallicities 
is reported in Table 1.

\begin{table*}
\begin{center}
\caption{Main properties of the AGB models of different metallicity (reported in col. 1)
used in the present analysis. Col. 2-6 report the initial abundances of various
elements, with the usual scale (X/H)$=12+\log[n(X)/n($H$)]$. Col. 7-9 report,
respectively, the range of masses used, the minimum mass undergoing HBB and the minimum
mass reaching the C-star stage (all these quantities are given in solar units).}                                       
\begin{tabular}{c c c c c c c c c}        
\hline\hline                        
Z  &   (He/H)  &   (C/H)  &   (N/H)  &   (O/H)  &   (Ne/H) & $\Delta M$ & 
$M_{HBB}^{min}$ &  $M_{C-star}^{min}$ \\
\hline       
$10^{-3}$ & 10.92 & 6.975 & 6.376 & 7.685 & 6.932 & 1-7.5 & 3  & 1.25       \\
$2\times 10^{-3}$ & 10.92 & 7.276 & 6.677 & 7.986 & 7.233 & 1-7.5 & 3 & 1.25 \\
$4\times 10^{-3}$ & 10.95 & 7.746 & 7.147 & 8.266 & 7.502 & 1-8 & 3.5 & 1.25 \\
$8\times 10^{-3}$ & 10.95 & 8.05 & 7.45 & 8.56 & 7.806 & 1-8 & 3.5 & 1.25 \\
$0.014$ & 10.97 & 8.432 & 7.871 & 8.731 & 7.915 & 1-8 & 3.5 & 1.5 \\
$0.018$ & 11.00 & 8.568 & 7.965 & 8.875 & 8.129 & 1-8 & 3.5 & 1.5 \\
$0.04$ & 11.06 & 8.941 & 8.339 & 9.249 & 8.503 & 1-8 & 4 & - \\
 \hline      
\end{tabular}
\end{center}
\end{table*}

\subsection{Dust formation}
Dust formation in the winds of AGB stars is described according to the schematization
introduced by \citet{fg06}. The wind is assumed to expand 
isotropically under the effects of radiation pressure, acting on dust grains, partly
counterbalanced by gravity. The dynamics of the wind is described by means of the
momentum conservation equation and by mass conservation, giving the radial
stratification of density as a function of the gas velocity and of the rate of
mass loss.

The effects of the radiation pressure is calculated by means of the opacity coefficient,
which, in turn, depends on the number and the size of the dust particles formed.
The growth rate of the dust particles of a given species are found via the 
difference between the growth and the vaporisation terms.

All the relevant equations, with an exhaustive discussion on the role played by various
physical factors, are given in \citet{fg06}.

The dust species considered depend on whether the surface of the star is oxygen-rich
or carbon rich: in the former case we consider the formation of silicates and of alumina
dust, whereas for carbon stars we model the formation and growth of silicon carbide
and of solid carbon grains \citep{ventura12a, ventura12b}.

\section{Changes in the surface chemistry of AGB stars}

The AGB models used in the present analysis were introduced and discussed in previous
papers by our group. We address the interested reader to \citet{ventura14b} 
($Z=4\times 10^{-3}$), \citet{ventura13} ($Z=1,8\times 10^{-3}$, initial mass above 
$3~M_{\odot}$), \citet{ventura14a} (low--mass models of metallicity $Z=1,8\times 10^{-3}$ 
and initial mass below $3~M_{\odot}$), paper II ($Z=2\times 10^{-3}$) and 
\citet{marcella16} ($Z=0.018$). To complete the array of comparison models we also 
introduce here a series of updated, unpublished models with {\it solar} metallicity 
$Z=0.014$ and with $Z=0.04$. 

\subsection{Low mass domain: the formation of carbon stars}
The surface chemistry of stars of mass below $\sim 3~M_{\odot}$ 
is altered only by the first dredge up (FDU) and by a series of TDU events, which
may eventually turn the star into a carbon star. The number of TDU experienced is higher the
larger is the initial mass, as more massive objects start the AGB phase with a more massive
envelope: this is the reason why only stars with initial mass above a threshold value will
eventually become carbon stars. The minimum mass required
to become carbon star, shown in Table 1 (col. 9) depends on the metallicity: the 
higher is Z, the more difficult is
to achieve C/O ratios above unity, owing to the larger quantity of oxygen in the star.
For sub-solar metallicities, the lowest mass becoming carbon star is $\sim 1.25~M_{\odot}$; 
in the solar case this lower limit is $\sim 1.5~M_{\odot}$, whereas no carbon stars
are expected to form for $Z=0.04$. The upper limit in mass for carbon stars coincides with 
the minimum mass required to ignite HBB; the latter process prevents the achievement
of the C-star stage, via destruction of the surface carbon. 

The chemical composition of carbon stars will be enriched in nitrogen, as a consequence
of the FDU. An increase in the surface oxygen, significantly smaller in comparison to 
carbon, is also expected, particularly in low-metallicity stars.

\subsection{Hot bottom burning and helium enrichment in massive AGB stars}
Stars with with initial mass above a given threshold
experience HBB during the AGB phase. The minimum mass required for the ignition
of HBB depends on the metallicity. Table 1 (col. 8) reports the values 
corresponding to the different metallicities.

The ignition of HBB strongly affects the AGB evolution, because the proton-capture 
nucleosynthesis activated at the base of the convective envelope significantly changes the 
surface chemical abundances of these stars. Among all, it provokes the destruction of the 
surface carbon and the production of great quantities of nitrogen. While this is a common 
property of all $M \geq 3.5~M_{\odot}$ models, the destruction of the surface oxygen, which
requires higher HBB temperatures ($\sim 80$~MK), is sensitive to the metallicity, and 
is higher the lower is $Z$ \citep{ventura13}. The destruction of oxygen is extremely 
sensitive to the modelling of convection: in the present analysis, based on the FST 
description, we find significant depletion of oxygen in metal poor AGB stars; conversely, 
when a less efficient convective model is used, the HBB experienced is weaker, thus 
limiting the efficiency of oxygen burning \citep{vd05, marcella16}. In this context, the
detection of oxygen-poor PNe, enriched in nitrogen, would be an important evidence 
in favour of a very efficient convective transport of energy in the internal regions
of the envelope of AGB stars.

The stars experiencing HBB are also exposed to the second dregde-up (SDU), after the 
consumption of the helium in the core. The main effect of the SDU is the
increase in the surface helium ($\Delta Y$), which is sensitive to the initial mass of the
star: typically, $\Delta Y$ is negligible in stars with mass
close to the minimum threshold required to start HBB, and increases with the initial
mass, up to $\Delta Y \sim 0.1$ for $M = 8~M_{\odot}$. This result is much more robust 
than the predictions concerning the depletion of oxygen, because the SDU takes place 
before the TP phase, thus the results are unaffected by most of the uncertainties
affecting AGB modelling.

\subsection{Surface chemistry at the end of AGB evolution}
\label{finchem}
\begin{figure*}
\begin{minipage}{0.495\textwidth}
\resizebox{1.\hsize}{!}{\includegraphics{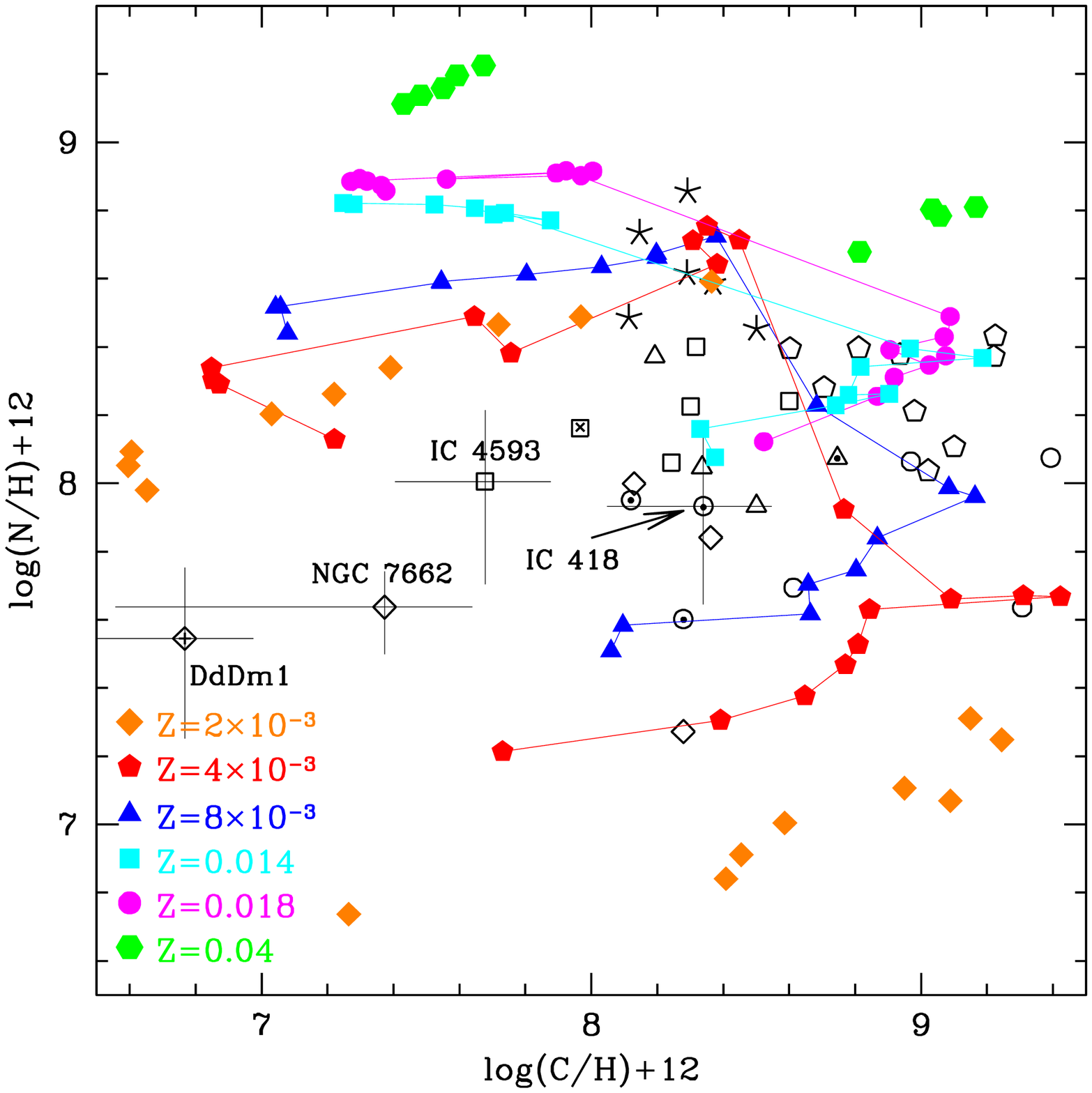}}
\end{minipage}
\begin{minipage}{0.495\textwidth}
\resizebox{1.\hsize}{!}{\includegraphics{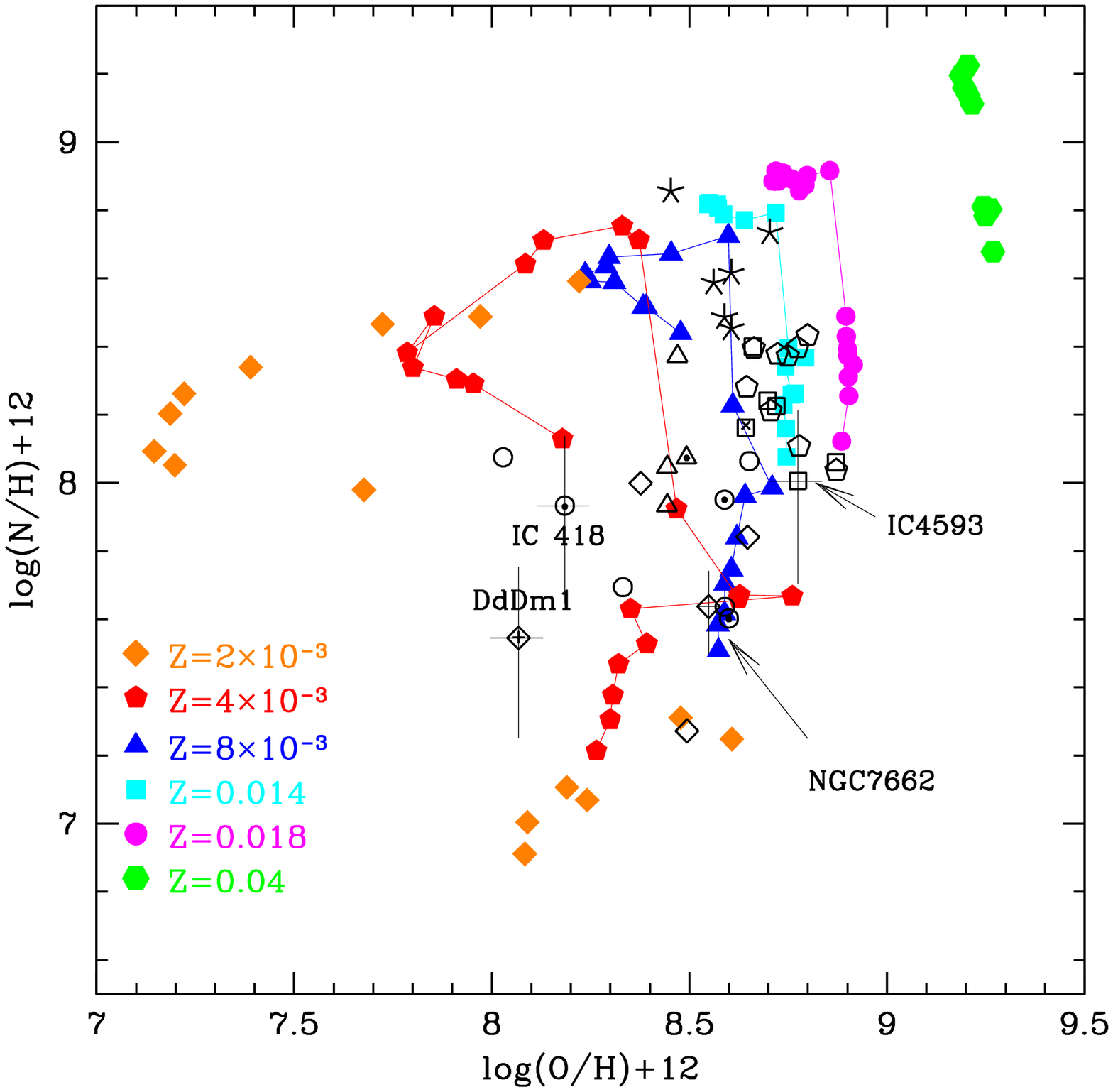}}
\end{minipage}
\vskip-00pt
\caption{The yields resulting from our AGB models in the {\it CN} (left) and {\it ON} 
(right) planes, superimposed with measured PN abundances of Sample 1.
The model sequences for each initial metallicity mixture are connected with solid lines,
with the exception of the $Z=2\times 10^{-3}$ and $Z=0.04$, which were not connected
for clarity reasons.
Each model point indicates the yields for a specific initial metallicity, and for a stellar 
mass. The individual sequences span the metallicity range given by the figure legend, and  
are indicated with different symbols and colors, $Z=2\times 10^{-3}$: orange
diamonds; $Z=4\times 10^{-3}$: red pentagons; $Z=8\times 10^{-3}$: blue triangles;
$Z=0.014$: cyan squares; $Z=0.018$: magenta circles; $Z=0.04$: green hexagons.
The chemical composition of the {\it Sample 1} PNe observed are indicated with black symbols, 
distinguished according to our interpretation in the text: open squares indicated the 
progeny of solar metallicity, low-mass AGB stars, that failed to reach the C-star stage; 
open diamonds indicate similar mass progenitors, but with sub-solar metallicity; open 
pentagons and circles both indicate progenitors with initial mass in the 
$1.5~M_{\odot} < M < 3~M_{\odot}$ range that go through the carbon stars phase, with 
metallicity respectively solar and sub-solar; open triangles (sub-solar metallicity) and 
asterisks (solar chemistry) are PNe whose progenitor AGB stars experienced HBB. PNe with 
evidence of carbon dust in their spectra are indicated with a dot inside the corresponding 
symbol, whereas PNe with silicates dust are shown with a cross inside the symbol. 
For the outliers discussed in section \ref{outliers1} we show the error bars for each data point.
}
\label{fpne1}
\end{figure*}

The final chemical composition of the models used here, which will be used to interpret
the PN abundances, are shown in Fig.~\ref{fpne1}, in the CN (left) and ON (right) planes. 
The distribution of the mass fractions of the individual species at the end of the AGB 
evolution models in these planes allows us to understand the role played by mass and 
metallicity on the evolution of the chemistry of the surface layers in AGB stars. 
For all the metallicities investigated, the lines connecting models of different mass 
define a typical counterclockwise shape, moving from the lowest ($1~M_{\odot}$) to the 
highest $8~M_{\odot}$ mass stars considered. As shown in the left panel of Fig.~\ref{fpne1},
the range of carbon abundances spanned by the models extends over two order of magnitudes, 
independently of the metallicity. Conversely (see right panel of Fig.~\ref{fpne1}), 
the distribution of the oxygen abundances is much more sensitive to metallicity: $Z=0.04$ models 
exhibit a negligible variation in oxygen, whereas in the $Z=2\times 10^{-3}$ case we find
an overall variation of a factor $\sim 30$. This behaviour is due to the larger
sensitivity of C to HBB and TDU, compared to O. TDU favours a significant increase in the
surface C, whereas the effects on O are much smaller. The activation of HBB provokes the 
destruction of the surface C, independently of Z, whereas the destruction of the surface 
oxygen via HBB is limited to the stars with the lowest metallicity. 

The lowest masses considered never become carbon stars, because they loose the 
external mantle before the surface carbon
exceeds the oxygen content. Compared to the initial chemical composition, with which they
formed, their chemistry is enriched in nitrogen, as a consequence of the FDU. For carbon 
the situation is more tricky. If no TDU occurred, the final carbon is smaller than
the initial quantity, because during the FDU the surface convection reached regions
where carbon was consumed by CN nucleosynthesis; however, if some TDU events take place,
the situation is reversed, and the final carbon is above the starting abundance.
We remark here that for these stars, owing to the null effects of HBB and the small
effects of TDU, the final chemical composition is extremely dependent on the assumptions
regarding the chemical mixture of the gas from which the stars formed.

As more massive objects, up to $\sim 3~M_{\odot}$, are considered, the 
theoretical sequences move to the right on the CN plane, owing to the increase in the 
surface carbon, as a consequence of repeated TDU events. On the contrary, nitrogen keeps 
approximately constant in this range of mass. The final carbon, could be potentially used 
as a mass indicator for $M < 3~M_{\odot}$ stars, because stars of higher mass reach higher 
carbon abundances in the final AGB phases.

A word of caution on the predictions regarding the evolution of the surface nitrogen of
$M \leq 3~M_{\odot}$ stars is needed here. In the present models, when modelling the
red giant branch (RGB) phase, neither thermohaline nor any sort of extra-mixing was
considered; this is going to underestimate the N increase, which occurs during
the ascending of the RGB \citep{charbonnel10}. Therefore, the N abundances of the models 
of initial mass below $\sim 3~M_{\odot}$ are to be considered as lower limits, with an 
overall uncertainty of $\sim 0.2$ dex. 

Regarding the possibility of using the abundances of some chemical elements to
infer the metallicity of the progenitors of the PNe, it is evident from the models locii 
of Fig.~\ref{fpne1} (left panel), that carbon should not be used to infer metallicity of 
the PN progenitors, since the abundance of this element is determined by the number of 
TDU events experienced by the star, which is scarcely related to metallicity. 
PN oxygen abundances, on the other hand, can be used as probes of the progenitor's 
metallicity, at least in the higher metallicity domain. In fact, in the right panel 
of Fig.~\ref{fpne1}, we observe a tight correlation between metallicity and oxygen 
abundances, for $Z>4\times 10^{-3}$. For lower metallicities, we should also consider: 
a) for massive progenitors, HBB in low-Z AGB stars favours 
a significant decrease in the surface O, which breaks out any O-Z relationship; b) in
low-mass progenitors the surface oxygen increases under the effects of TDU (see the
$Z = 4\times 10^{-3}$ line in the figure). 

\citet{garcia16a} interpreted oxygen enhancement observed in low-metallicity Galactic 
PNe based on these same modelling feature.

The nitrogen abundance is also correlated to metallicity in the low-mass domain; 
however, the afore mentioned uncertainties affecting the extent of the nitrogen increase 
during the RGB phase prevents the use of the measured N as a robust metallicity indicator.

Stars with mass above the minimum threshold required to activate HBB, 
$M > 3-3.5~M_{\odot}$, show the imprinting of CN or, in some cases, CNO
nucleosynthesis, in their surface chemical composition. A robust prediction in this
case is the significant increase in the nitrogen content, which at the end of the AGB
phase is a factor of $\sim 10$ higher than the N initially present in the star. 
The final carbon of these stars is more uncertain, as it is sensitive
to the relative importance of HBB and TDU. In an HBB-dominated environment the final C 
will be almost a factor $\sim 10$ smaller than the initial value; however, a few TDU 
events, in the very final AGB phases, after HBB was turned off, might partly 
counterbalance the effects of the proton-capture nucleosynthesis; this 
argument is still debated. According to the models used here, shown in the left panel
of Fig.~\ref{fpne1}, we find that stars of initial mass $\sim 4-5~M_{\odot}$
evolve to final phases characterised by a great increase in N and a carbon content
similar to the matter from which they formed; this is because in these stars a few final
TDU episodes make the surface carbon, previously destroyed by HBB, to rise again; 
conversely, $6-8~M_{\odot}$ stars reach the final evolutionary stages with a surface 
carbon reduced by almost an order of magnitude in comparison with the initial chemistry.
The effect of metallicity in this region of the CN plane is that higher Z stars evolve
to higher N: this is because the equilibrium abundance of N in a region where CNO
nucleosynthesis is active is proportional to the overall C+N+O content.

Concerning oxygen, the results are metallicity-dependent: as shown in the right
panel of Fig.~\ref{fpne1}, the extent of the depletion of oxygen in massive AGB
stars is negligible in solar chemistry models, whereas it amounts to almost a
factor 10 in the $Z=2\times 10^{-3}$ case.

\section{Comparison of the observed PN abundances with the AGB final yields}
Our aim is to study a large sample of Galactic PNe, for which the chemical abundances
were derived homogeneously, in the framework of AGB evolution.

To trace AGB evolution and especially the HBB and TDU phenomena, the most important 
abundances are those of carbon, oxygen, and nitrogen. Dust properties of the PNe add 
another handle to determine carbon enrichment (see Stanghellini et al. 2012) even in the 
cases where carbon abundances are not available.

Stanghellini et al. (2007) found clear correlations between gas and dust composition
for a sample of PNe in the LMC.
PNe with carbon-rich dust (CRD) features were found to have typically carbon-rich gas as well
(i.e. $C/O > 1$), while PNe with oxygen-rich dust (ORD) features had $C/O < 1$. The same
analysis could not be done for Galactic PNe (Stanghellini et al. 2012) for the 
unavailability of dust and gas chemistry in the same sample of Galactic PNe.

It is worth recalling here that oxygen and nitrogen abundances in Galactic PNe are within 
easy reach from optical spectra. On the other hand, carbon abundances, and information on 
the nature of the nebular dust, are observationally harder to determine.
Carbon in PNe can be measured via collissionally-excited emission lines (CELs), the 
majority of which are emitted in the UV regime, with C II] at 2626-28 \AA, and C III] at 
1907-09 \AA. These two intensities are usually sufficient for a complete carbon 
determination for low and intermediate excitation PNe. For high excitation PNe, the UV 
recombination line C IV 1548-50 \AA~ is also used. 

An additional possibility is to estimate the C/O ratio from optical recombination 
lines (ORLs), which are much easier to acquire. There are several C/O estimates in the 
literature from OLRs analysis, both from the assumption that C$^{2+}$/O$^{2+} \sim$ C/O, 
and from ICF evaluation (see for a discussion 
Delgado-Inglada \& Rodr{\'{\i}}guez 2014). 
However, we prefer to avoid carbon estimates from ORLs in this study, to avoid mixing 
determinations from CELs and RLs.

The dust content of PNe has been studied in recent years with the advent of the Spitzer 
Space Telescope. IRS/Spitzer spectra have been exploited to determine whether PNe have 
carbon or oxygen rich dust, or a mixture of both (see Stanghellini et al. 2012; 
Garc{\'{\i}}a-Hern{\'a}ndez \& G{\'o}rny 2014). The PNe whose dust content has been 
classified into 
carbon-rich and oxygen-rich classes to date do not overlap with those with a CELs carbon 
determination. For this reason, we decided to select two PN samples for the AGB model 
comparison: the first ({\it Sample 1}) is driven by the availability of carbon abundances 
determined from CELs in the literature; the second ({\it Sample 2}) is driven by being 
classified based on their dust contents based on IRS/Spitzer data.

\subsection{{\it Sample 1} PNe}
{\it Sample 1} PNe are Galactic PNe whose carbon, oxygen, and nitrogen abundances are 
available in the literature to date, with carbon abundances determined from UV emission 
lines.  
Before the HST had became available, UV spectra of Galactic PNe have been acquired 
with the IUE satellite. Data have been accumulated in the decades, and two main groups 
have revisited the IUE spectra and derived carbon abundances of Galactic PNe. Kingsburg 
and Barlow (1994), and Henry et al. (2000) published  carbon abundances, with one target 
in common. We also searched the literature for Galactic PNe whose spectra has been acquired 
with the HST. Henry et al. (2015) and Dufour et al. (2015) examined 7 carbon determinations 
in Galactic PNe (4 in common with the IUE samples described above). Furthermore, Henry et 
al. (2008) observed the halo PN DdDm~1 (PN~G061.9+41.3), and Bianchi et al. (2001) observed 
a globular cluster PN (K648 in M15). 

In summary, reliable carbon abundances are available for 40 Galactic PNe from UV data, 7 
of which are from HST spectra. For all these PNe, the original papers also gave abundances 
of He, N, O, Ne. We list {\it Sample 1} PNe in Table 2, where we give their PN~G numbers 
(column 1), their usual names (column 2), their dust and morphological types (columns 3 
and 4, see table note for description of the keys), and their C, N, and O abundances in 
the usual scale of log(X/H)+12 (columns 5 through 7). The asymmetric log uncertainties 
have been calculated from the uncertainties in the original references, when given. The 
references for the abundances are given in column 8.

It is worth noting that all abundance references use the same ICF scheme 
(Kingsburgh \& Barlow 1994), and are thus homogeneous, with the exception of the 7 PNe 
whose abundances are from Dufour et al. (2015). We comment on possible use of the model 
abundances from Henry et al. (2015), based on Dufour et al.' s (2015) data, in specific 
cases in the following sections.
It is important to note that we have checked all PNe in this sample for possibly being 
located in the bulge or halo of the Galaxy, according to the definition which is commonly 
adopted (see Stanghellini \& Haywood 2010). We found none of the {\it Sample 1} PNe to 
belong to the bulge, while H~4-1 (PN~G049.3+88.1), BoBn~1 (PN~G108.4-76.1), DdDm~1 
(PN~G061.9+41.3), and Me~2-1 (PN~G342.1+27.5) may belong to the Galactic halo.

\begin{table*}
\caption{Chemical abundances and other properties of {\it Sample 1} PNe}                                       
\begin{tabular}{l l r r r  r r  l}        
\hline\hline   
 PN~G& Name&  Dust&   Morph.&   log(C/H)+12& log(N/H)+12& log(O/H)+12& Ref.\\ 
 (1)&(2)&(3)&(4)&(5)&(6)&(7)&(8)\\
 \hline \\
 025.3+40.8&   IC~4593&$\dots$&2&7.678$^{+0.199}_{-0.275}$& 8.005$^{+0.210}_{-0.301}$&8.775$^{+0.061}_{-0.070}$&1  \\
036.1--57.1&  NGC~7293&$\dots$&2&8.602$^{+0.073}_{-0.082}$& 8.395$^{+0.117}_{-0.148}$&8.663$^{+0.017}_{-0.017}$&1  \\
037.7--34.5&  NGC~7009&$\dots$&4&8.318$^{+0.205}_{-0.289}$& 8.400$^{+0.208}_{-0.295}$&8.661$^{+0.061}_{-0.071}$&2  \\
043.1+37.7&  NGC~6210& 4&3&7.966$^{+0.201}_{-0.279}$& 8.162$^{+0.205}_{-0.288}$&8.643$^{+0.061}_{-0.071}$&1  \\
049.3+88.1&     H~4-1&$\dots$&$\dots$&8.613$^{+0.100}_{-0.114}$& 7.694$^{+0.092}_{-0.103}$&8.332$^{+0.039}_{-0.042}$&1  \\
051.4+09.6&    Hu~2-1& 3&4&8.746$^{+0.188}_{-0.269}$& 8.073$^{+0.085}_{-0.094}$&8.493$^{+0.041}_{-0.046}$&1  \\
061.4--09.5&  NGC~6905&$\dots$&2&8.243 & 8.061 &8.871 &2  \\
061.9+41.3&    DdDm~1& 5&$\dots$&6.767$^{+0.208}_{-0.294}$& 7.545$^{+0.208}_{-0.294}$&8.068$^{+0.062}_{-0.073}$&1  \\
063.1+13.9&  NGC~6720&$\dots$&2&8.811$^{+0.208}_{-0.295}$& 8.397$^{+0.208}_{-0.296}$&8.774$^{+0.061}_{-0.071}$&1  \\
083.5+12.7&  NGC~6826&2& 2&8.120$^{+0.210}_{-0.301}$& 7.951$^{+0.204}_{-0.286}$&8.589$^{+0.060}_{-0.070}$&1  \\
084.9--03.4&  NGC~7027&$\dots$&4&8.980$^{+0.141}_{-0.168}$& 8.211$^{+0.124}_{-0.144}$&8.706$^{+0.061}_{-0.070}$&1  \\
106.5--17.6&  NGC~7662&$\dots$&2&7.373$^{+0.267}_{-0.817}$& 7.637$^{+0.104}_{-0.138}$&8.549$^{+0.027}_{-0.029}$&3  \\
108.4--76.1&    BoBn~1&$\dots$&$\dots$&9.393$^{+0.239}_{-0.384}$& 8.075$^{+0.119}_{-0.138}$&8.029$^{+0.057}_{-0.066}$&1  \\
123.6+34.5&   IC~3568&$\dots$&1&8.193$^{+0.067}_{-0.079}$& 8.372$^{+0.207}_{-0.292}$&8.470$^{+0.030}_{-0.032}$&3  \\
130.9--10.5&   NGC~650&$\dots$&4&9.220$^{+0.237}_{-0.372}$& 8.372$^{+0.119}_{-0.137}$&8.749$^{+0.061}_{-0.070}$&1  \\
161.2--14.8&   IC~2003&$\dots$&3&8.969 & 8.064 &8.652 &2  \\
189.1+19.8&  NGC~2372&$\dots$&4&8.301 & 8.225 &8.720 &2  \\
194.2+02.5&     J~900&$\dots$&3&9.307 & 7.636 &8.590 &2  \\
197.8+17.3&  NGC~2392&$\dots$&2&8.336$^{+0.206}_{-0.291}$& 8.046$^{+0.207}_{-0.293}$&8.444$^{+0.061}_{-0.071}$&1  \\
206.4--40.5&  NGC~1535&$\dots$&2&8.279$^{+0.102}_{-0.116}$& 7.272$^{+0.128}_{-0.150}$&8.494$^{+0.061}_{-0.071}$&1  \\
215.2--24.2&    IC~418& 3&2&8.340$^{+0.207}_{-0.292}$& 7.933$^{+0.205}_{-0.288}$&8.185$^{+0.061}_{-0.071}$&1  \\
221.3--12.3&   IC~2165&$\dots$&2&8.501$^{+0.024}_{-0.025}$& 7.933$^{+0.027}_{-0.028}$&8.444$^{+0.026}_{-0.027}$&3  \\
231.8+04.1&  NGC~2438&$\dots$&2&8.130 & 7.999 &8.377 &2  \\
234.8+02.4&  NGC~2440&$\dots$&4&8.290$^{+0.036}_{-0.040}$& 8.616$^{+0.027}_{-0.028}$&8.606$^{+0.051}_{-0.057}$&3  \\
243.3--01.0&  NGC~2452&$\dots$&2&8.501 & 8.453 &8.606 &2  \\
261.0+32.0&  NGC~3242&2&2&8.279$^{+0.018}_{-0.019}$& 7.602$^{+0.105}_{-0.138}$&8.600$^{+0.011}_{-0.011}$&3  \\
265.7+04.1&  NGC~2792&$\dots$&2&8.924 &  $\dots$ &8.759 &2  \\
278.1--05.9&  NGC~2867&$\dots$&2&9.100 & 8.107 &8.778 &2  \\
278.8+04.9&      PB~6&$\dots$&$\dots$&9.225 & 8.432 &8.799 &1  \\
285.7--14.9&   IC~2448&$\dots$&2&8.936 & 8.378 &8.723 &2  \\
296.6--20.0&  NGC~3195&$\dots$&3&8.600 & 8.241 &8.698 &2  \\
307.2--03.4&  NGC~5189&$\dots$&4&8.114 & 8.486 &8.589 &2  \\
309.1--04.3&  NGC~5315&$\dots$&4&8.369$^{+0.047}_{-0.053}$& 8.587$^{+0.033}_{-0.035}$&8.562$^{+0.021}_{-0.022}$&3  \\
312.3+10.5&  NGC~5307&$\dots$&3&8.210 &  $\dots$ &8.629 &2  \\
315.0--00.3&  He~2-111&$\dots$&4&8.292 & 8.856 &8.453 &2  \\
320.3--28.8&  He~2-434&$\dots$&3&8.362 & 7.841 &8.648 &2  \\
320.3--28.8&  NGC~5979&$\dots$&2&8.706 & 8.281 &8.752 &2  \\
327.8+10.0&  NGC~5882&$\dots$&2&7.895$^{+0.101}_{-0.132}$& 7.760$^{+0.091}_{-0.116}$&8.645$^{+0.031}_{-0.034}$&3  \\
341.8+05.4&  NGC~6153&$\dots$&2&8.146 & 8.735 &8.704 &2  \\
342.1+27.5&    Me~2-1&$\dots$&2&9.021 & 8.037 &8.872 &2  \\
  \hline

\end{tabular}

{Dust type is either featureless (F, 0); carbon rich dust (CRD) aromatic (1); CRD aliphatic 
(2); CRD both aromatic and aliphatic (3);  oxygen rich dust (ORD) crystalline (4); ORD 
amorphous (5);  ORD both crystalline and amorphous (6);  mixed chemistry dust (MCD, 7).
Morphology is  Round (1);  Elliptical (2)  Bipolar Core (3);  Bipolar (4);  Point-symmetric (5).
Elemental abundance references  are from a series of papers summarized in Henry et al. (2000, 1); 
from  Kingsburg \& Barlow (1994, 2); from Dufour et al. (2015, 3).}
\end{table*}

\subsubsection{The origin of the {\it Sample 1} PNe} 

We can take full advantage of the availability of gaseous carbon abundances in {\it sample 1} 
PNe to interpret them by comparison with the AGB yields. 
Combined with the measurements of nitrogen and oxygen, this allows the knowledge of the
overall CNO chemistry, which can be used to infer the progenitors of the individual
sources in the sample. We therefore followed an approach similar to paper I and paper II

Fig.~\ref{fpne1} shows the {\it Sample 1} PN chemical abundances in the CN
(left panel) and ON (right panel) planes. Superimposed to the data we show the final yields 
of AGB models of various mass and metallicity, discussed in the previous
section.
The symbols used to indicate the PNe reflect our understanding of the mass and
chemical composition of the progenitors. We reiterate here that for those cases
when the N and O abundances observed provided different directions for interpretation, we 
relied on the O determinations, because of the on-the-average smaller errors 
associated to the measurements of oxygen compared to nitrogen, and for the uncertainty
affecting the predictions of the variation of the surface nitrogen in low-mass AGB stars, 
owing to the still debated effects of extra-mixing during the RGB ascending.

Approximately $65\%$ of {\it Sample 1} PNe descend from solar metallicity 
stars, whereas $\sim 35\%$ have a slightly sub-solar
chemistry, with metallicity $Z\sim 4-8\times 10^{-3}$.

About half of {\it Sample 1} PNe descend from carbon stars. Their surface chemical
composition was modified mainly by TDU, with no effects of HBB. These sources are 
indicated in Fig.~\ref{fpne1} with open pentagons (solar metallicity) and circles 
(sub-solar chemistry). The progenitors of this group of PNe, characterised by masses in 
the range $1.5~M_{\odot} < M < 3~M_{\odot}$, formed between 2 Gyr and 500 Myr ago. 
According to our interpretation, the PNe belonging to this group 
with the largest carbon abundance are younger and 
descend from higher mass progenitors. In this sub-sample we find NGC~3242 (PN~G261.0+32.0), 
NGC~6826 (PN~G083.5+12.7), and IC~418 (PN~G215.2-24.2), whose IR spectra, analysed by 
\citet{delgado15}, exhibit traces of carbon dust, consistently with the interpretation 
given in \citet{garcia16a} and confirmed in the present study.

$30\%$ of the PNe in this sample descend from low-mass progenitors, with mass in the  
$\sim 1-1.5~M_{\odot}$ range, that never reached the C-star stage. These stars, indicated 
with open squares (solar chemistry) and diamonds (sub-solar metallicity) in Fig.~\ref{fpne1}, 
are the oldest PNe, formed between 10 Gyr and 2 Gyr ago. This group of PNe includes NGC 
6210 (PN~G043.1+37.7), also present in the sample studied by \citet{delgado15}, and 
interpreted by \citet{garcia16b} as the progeny of a low-mass progenitor.

This sample also include a group of objects that experienced HBB. These PNe are 
indicated as asterisks and open triangles in Fig.~\ref{fpne1}, according to whether
their metallicity is, respectively, solar, or sub-solar. We used once more the 
combination of the O and N abundances to deduce the metallicity. These
PNe descend from stars of mass above $3.5~M_{\odot}$, formed in more recent epochs,
younger than 250 Myr. Their surface chemistry, largely contaminated by HBB, exhibits
extremely large N abundances; the carbon of these PNe, in all cases above 8, suggests 
the additional effects of TDU and seems to rule out $7-8~M_{\odot}$ progenitors. The 
helium abundances of these sources, in all but one case $\log(He/H)+12 > 11.05$, 
further supports this interpretation. In 
this group we include Hu2-1 (PN~G051.4+09.6), surrounded by carbon dust, suggesting 
the combined effects of HBB and TDU.

\subsubsection{Similarities and differences of {\it Sample 1} Galactic PNe with the PNe 
in the Magellanic Clouds}

The PNe of {\it Sample 1} are the only Galactic PNe with measured carbon from CELs. 
There are two other notabe samples of PNe with measured carbon abundances from CELs, 
namely, those in the SMC and the LMC, discussed, respectively, in paper I and paper II. 
We can now compare results from the three different galaxies, based on carbon abundances.

According to the results found in this paper, we confirm one of the main findings of 
papers I, and II: the present models of carbon stars nicely reproduce the largest 
abundances of gaseous carbon observed. The observations indicate $\log(C/H)+12<9.2$ across 
the galaxies studied, in agreement with the models. This confirms the existence of an upper 
limit to the amount of carbon which can be accumulated in the external regions of AGB stars. 
This limit is likely due to the formation of large quantities of carbon dust in the winds 
of carbon stars, which favours a fast loss of the external mantle, owing to the effects of 
radiation pressure, acting on dust grains.

In {\it Sample 1} PNe we did not find pure HBB contamination, at odds with what found for 
the Magellanic Cloud PNe in papers I and II: while a few N-rich PNe in the LMC (see left 
panel of Fig~4 of paper I) and the SMC (see left panel of Fig~2 of paper II) disclosed 
extremely low carbon abundances ($\log(C/H)+12<8$), here we find $\log(C/H)+12>8.4$ for 
all N-rich PNe. The small number of N-rich PNe in all studied galaxies does not allow any 
robust statistics. However, part of the explanation of this result could reside in the 
averagely higher metallicities of {\it Sample 1} PNe compared with those of paper I and 
paper II, because the HBB experienced by massive AGB stars is stronger the lower is Z.

\begin{figure*}
\begin{minipage}{0.495\textwidth}
\resizebox{1.\hsize}{!}{\includegraphics{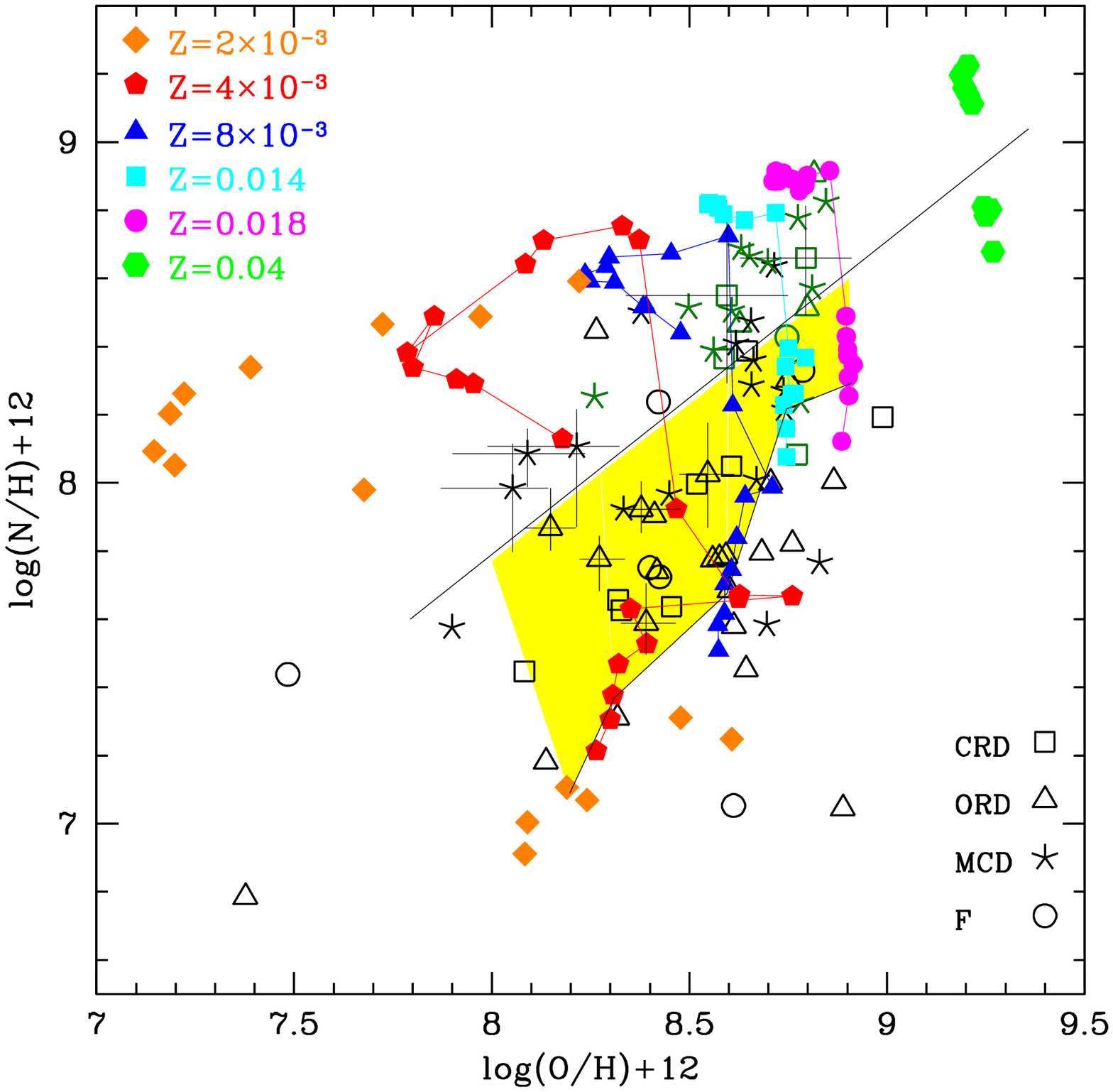}}
\end{minipage}
\begin{minipage}{0.495\textwidth}
\resizebox{1.\hsize}{!}{\includegraphics{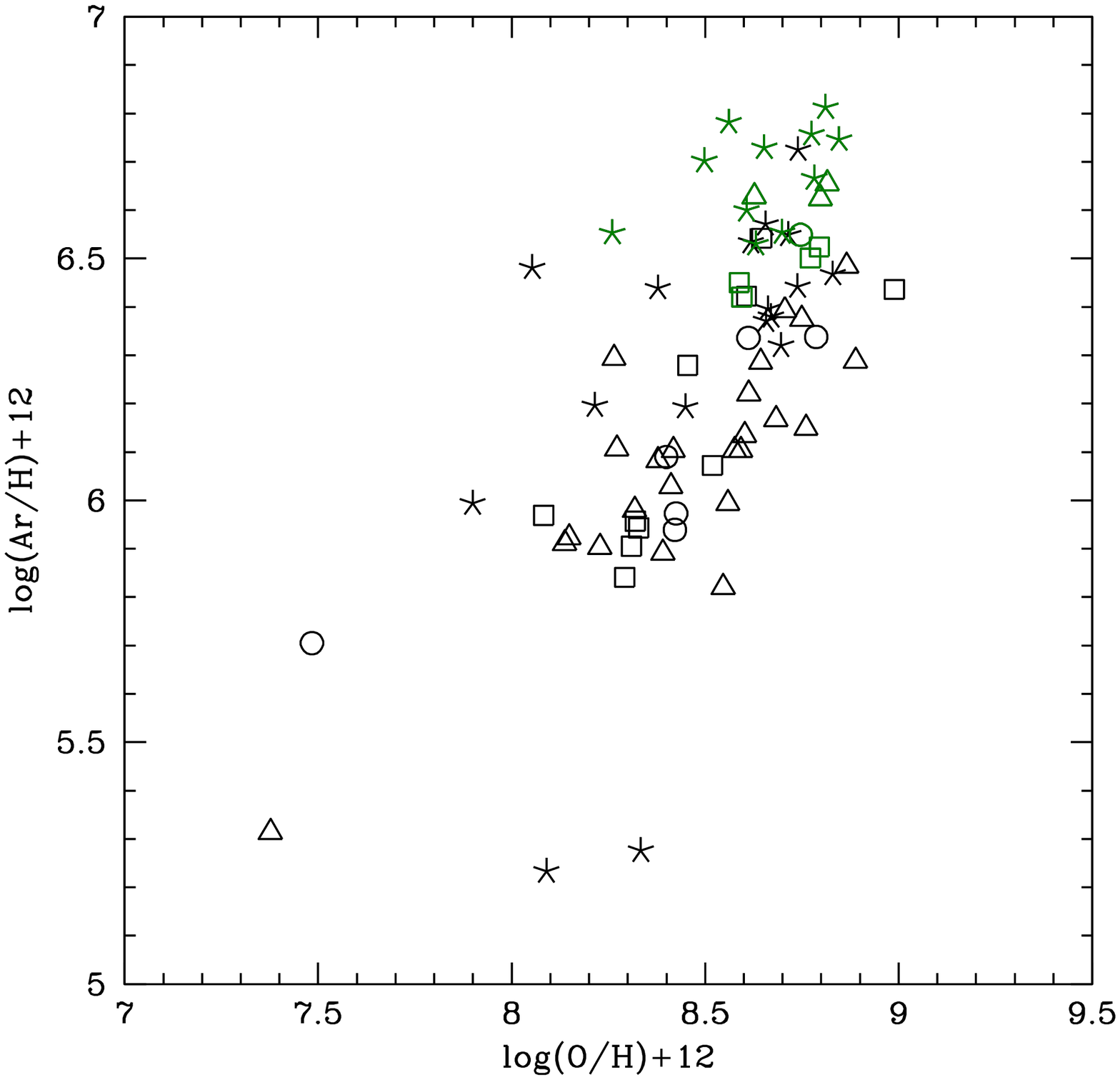}}
\end{minipage}
\vskip-00pt
\caption{Left: Yields from AGB modelling, in the ON plane, using 
the same symbols for the different metallicities as in Fig.~\ref{fpne1}. 
The distribution of the {\it Sample 2} PNe abundances are also shown here. The observed 
abundances are reported with black points, with different symbols, according to the dust 
features, as in the legend. For the outliers discussed in section \ref{outliers2} we
show the error bars. The diagonal line separates the region of the plane where we
expect to find the progeny of carbon stars (the shaded, yellow region) from the upper 
region, where we find the descendants of stars which experienced HBB.
Right: {\it Sample 2} PNe on the argon vs. oxygen plane.  
The PNe shown in green present evidence of helium enrichment, thus suggesting a massive
progenitor of mass above $\sim 4~M_{\odot}$, which experienced HBB during the AGB phase. }
\label{fgorny}
\end{figure*}

\subsubsection{A few outliers}
\label{outliers1}
In the analysis of the PNe in the present sample, as stated previously, we attempted
to deduce the main properties of the progenitors based on the combination of the CNO
abundances observed. While the agreement between the observations and the theoretical
expectations was generally extremely satisfactory, in a few cases we could not fit
simultaneously the abundances of all the elements. We analyse these PNe individually,
in the following.

{\it IC 4593 (PN~G025.3+40.8)}. The abundances of N, O and Ne suggest a low-mass ($\sim 1~M_{\odot}$)
progenitor, with solar metallicity. As evident in the left panel of Fig.~\ref{fpne1},
the only problem with this interpretation is the carbon content, with is $\sim 0.5$
lower than expected. A possible explanation could be that the
chemistry of IC 4593 reflects the sole effects of mixing during the RGB ascending, and
that some additional carbon depletion occurred, owing to unusually large
extra-mixing during the RGB phase. It is worth adding that both carbon and nitrogen 
error bars are very large for this PN and the inconsistency with the models could be 
ascribed to the low-quality data available.

{\it DdDm 1 (PN~G061.9+41.3)}. The O and N abundances indicate a low-mass, 
metal-poor chemistry, with $Z=2\times 10^{-3}$. The presence of traces of silicate
dust in the spectra is compatible with this hypothesis. While the carbon abundance
is substantially compatible with this interpretation, the measured N is a factor $\sim 4$
higher than expected (see the left panel of Fig.~\ref{fpne1}). Possible explanations
are an overestimation of the surface N and/or a difference in the original N content 
in comparison to the typical pop II chemistry, although the large errorbar for nitrogen 
may indicate a poor S/N spectrum. It is worth recalling that this is a halo PNe 
(Henry et al. 2008), thus the initial chemical mixture ratios used for the AGB models may 
not be the ideal choice to model it. 

{\it NGC 7662 (PN~G106.5-17.6)}. The N, O and Ne abundances indicate a low-mass progenitor, 
of sub-solar metallicity, $Z=8\times 10^{-3}$. As shown in the left panel of Fig.~\ref{fpne1},
the surface carbon (7.37) is too small for this interpretation (but note the huge errorbar). 
Note that the carbon abundance whose derivation is based on photoionisation models
${C/H}=8.13$ (Henry et al. 2015) is in much better agreement with our interpretation. 
NGC 7662 is a inhomogeneous PN, with a lot of 
stratification (see Dufour et al. 2015). This may help to explain the disagreement between 
the carbon and other abundance indicators. 

{\it IC 418 (PN~G215.2-24.2)}. The O and Ne abundances point in favour of a $\sim 2~M_{\odot}$ 
low metallicity progenitor, $Z\sim 4\times 10^{-3}$. This interpretation would also be in
agreement with the presence of carbon dust in the surroundings of this object. 
If this is the case, we deduce from the right panel of Fig.~\ref{fpne1} that the N 
content is overestimated by a factor of $\sim 2$. The carbon content is also an
issue in this case, as according to the models a higher amount of carbon is expected
(see left panel of Fig.~\ref{fpne1}). IC 418 was discussed in \citet{garcia16a}: the
interpretation was different in terms of metallicity, as the oxygen  content
derived by \citet{delgado15} is  significantly lower compared to the values upon which the 
present analysis is based.

\subsection{{\it Sample 2} PNe}

{\it Sample 2} PNe are those Galactic PNe whose dust spectrum has been observed by 
IRS/Spitzer, and whose dust properties have been uniformly analyzed in the recent past by 
Stanghellini et al. (2012), \citet{perea09} and Gutenkunst et al. (2008). 
Elemental abundances for these PNe are available from \citet{garcia14} (hereinafter GG14), 
who targeted explicitly these dust-analyzed PNe and recalculated the principal elemental 
abundances from published and newly observed optical emission lines.  As in turned out, 
unfortunately none of the {\it Sample 2} PNe have carbon abundance measured from the CELs 
UV lines, thus, while interesting, they lack one of the major observing constraints for 
this type of work. There is still purpose of using this large sample to confront with the 
AGB yields.

In Table 3 we give the {\it Sample 2} PN names, dust and morphological properties 
(note that the dust content codes and the morphological type codes are the same as in 
Table 2), and the elemental abundances (nitrogen, oxygen, and argon, see Fig.~\ref{fgorny}) 
with their uncertainties. It is worth noting that the {\it Sample 2} PNe, consisting of 101 
targets, is more restricted than the whole sample discussed by GG14. In fact, we have 
eliminated from the GG14 sample those PNe whose ionization correction factor were deemed 
by the authors to be uncertain (S.~K. G{\'o}rny, private communication), i.e., 
those with O$^{2+}/O<0.4$. In the Tables we mark with $^1$ and $^2$ the PNe that are 
likely to belong to the Galactic bulge or halo respectively, based on the prescription in 
Stanghellini \& Haywood (2010). 

We base our comparison between abundances and yields on the O-N plane. The N and O 
abundances observed are shown in the left panel of Fig.~\ref{fgorny}, overimposed to the 
results from AGB modelling. The observations have been indicated with different symbols, 
according to the dust properties. The yellow-shaded region indicates the zone of the O-N 
plane where we expect to find the progeny of carbon stars; in the interpretation of the 
PNe located close to the lower and upper borders of this region, we will consider the 
uncertainties related to the final nitrogen of low-mass stars, discussed in section 
\ref{finchem}. The right panel of Fig.~\ref{fgorny} shows the observed oxygen and
argon abundances. The PNe shown in green in both panels are those with helium
abundances $12+\log(He/H)>11.1$, which we will take as the typical threshold above
which we see the signature of the second dredge-up, operating in stars of initial
mass above $\sim 4~M_{\odot}$.

To derive the mass, age, and metallicity of the progenitors of the {\it Sample 2} PNe 
we rely on their position on the O-N plane to understand the relative importance of HBB
and TDU in modifying the surface chemical composition, which provides an indication
on the initial mass (see discussion in section \ref{finchem}). The metallicity of the
progenitors is deduced on the basis of the position on the O-Ar plane, shown in the right
panel of Fig.~\ref{fgorny}. Among the various species unaffected by AGB evolution, we
prefer to use Argon as metallicity indicator, because: a) the chlorine abundance is
available only for part of the PNe in the sample; b) the sulphur detected might not reflect 
the original content, because part of this element is absorbed in dust particles,
particularly in carbon-rich environments \citep{pottasch06}. 

\subsection{Mass and metallicity distribution}
The comparison between the {\it Sample 2} PN abundances and the models indicate that 
approximately half of the PNe in this sample have solar/supersolar metallicity, the remaining 
$\sim 50\%$ exhibit a sub-solar chemistry, with $Z_{\odot}/3 < Z < Z_{\odot}/2$. We find 
also a few metal-poor objects, namely M 2-39 (PN~G008.1-04.7), Pe 2-7 (PN~G285.4+02.2), 
M 4-6 (PN~G358.6+01.8): based on the N, O and Ar abundances, we interpret these PNe as the 
progeny of low-mass stars, with mass below $\sim 1.5~M_{\odot}$ and metallicity 
$Z=1-2 \times 10^{-3}$. The relative distribution of PNe of different metallicity 
exhibits a slight change according to the position in the Galaxy: in the bulge, the "solar"
component exceeds by $\sim 50\%$ the "sub-solar"group, whereas in the disk the solar
metallicity PNe account for $\sim 45 \%$ of the total, the more numerous component being 
the sub-solar one. 

The diagonal line in the left panel of Fig.~\ref{fgorny} represents an approximate 
separation between the stars that experienced
some HBB, with the consequent nitrogen enrichment of the surface layers, and those that
experienced only dredge-up effects. The PNe above this line are identified as the progeny
of stars of mass $M\geq 4~M_{\odot}$, which underwent SDU and HBB. The 
fact that most of the PNe in this zone of the O-N plane are also helium-rich adds more
robustness to this interpretation. These PNe are the objects formed more recently in
the sample examined, and are younger than $\sim 250$ Myr. The fraction of PNe which have 
been exposed to HBB during the AGB phase is approximately $30\%$. 

The PNe within the shaded region in the O-N plane are generally interpreted as the
progeny of stars with mass in the range $1.5~M_{\odot} < M < 3~M_{\odot}$, which reached 
the carbon star stage during the AGB evolution. This is the dominant component in the
sample, including $\sim 50\%$ of the PNe observed. Similarly to the analysis of 
{\it Sample 1} PNe, we may conclude that {\it Sample 2} PNe have ages in the range 
300 Myr - 2 Gyr, with no straightforward trend between age and the position on the O-N 
plane. We reiterate here that the vertical extension of this region is somewhat uncertain,
as it is sensitive to the extent of the N enrichment occurring during the RGB, which is
still debated, and may be different between stars of the same mass. 

The remaining $20\%$ of {\it Sample 2} PNe are located below the shaded region
in the left panel of Fig.~\ref{fgorny}; these are the descendants of low-mass stars,
which did not reach the C-star stage during the AGB phase. These are the oldest PNe,
formed between 2 Gyr and 10 Gyr ago. The spectra of the majority of the PNe in this 
sub-sample exhibit the feature of silicate dust, in agreement with our interpretation.
We do not expect any carbon enhancement in the surface chemical composition of these
objects.

\subsection{Galactic distribution of {\it Sample 2} PNe}
The mass and metallicity distribution of {\it Sample 2} PNe can be used to outline
some important points, regarding how PNe of different mass, chemistry and dust type are 
distributed across the Galaxy. 

According to our interpretation, PNe with solar/supersolar metallicity progenitors are 
composed by $\sim 50\%$ of objects that experienced HBB during their AGB life, an 
indication of massive and relatively young progenitors. The remaining half of solar 
metallicity PNe are divided among the progeny of C-stars and of low-mass stars, in 
approximately equivalent percentages. These relative numbers hold both for the bulge 
and the disk.

\subsection{A few outliers}
\label{outliers2}
The PNe Mac1-2 (PN~G309.5-02.9) and H1-33 (PN~G355.7-03.0, in the bulge), represented by 
the two open squares in the HBB region (i.e., above the straight line) in the left panel 
of Fig.~\ref{fgorny}, are enriched in nitrogen, thus indicating the signature of HBB. 
Their surface chemical composition suggests a $\sim 4-5~M_{\odot}$ progenitor, of solar 
metallicity. The problem with this interpretation is that their spectra exhibit the 
typical features of carbon dust; this is at odds with our understanding, as the stars 
that experience HBB during the AGB phase destroy the surface carbon, thus leaving no room 
for the formation of carbon dust in the circumstellar envelope. The only possibility to 
reconcile these results with our theoretical description is that the nitrogen of these 
PNe are overestimated (the errors associated to the N determination are of the order of 
0.2 dex), so that in the O-N plane they fall into the carbon star zone, yellow-shaded in 
Fig.~\ref{fgorny}. Alternatively, we are left with three possibilities: a) carbon dust can 
be formed around PNe, despite the surface oxygen is in excess of carbon; b) in the very 
final AGB phases, after HBB is turned off by the loss of the external mantle, a sequence 
of TDU events may favour the formation of a carbon star; (c) the PN is of the MCD type, 
but the oxygen dust features are too weak to be seen in the spectra. It goes without saying 
that the knowledge of the carbon content of Mac1-2 and H1-33 would be crucial to answer 
these questions.

The bulge PNe Th3-4 (PN~G354.5+03.3), M3-44 (PN~G359.3-01.8), and H1-61(PN~G006.5-03.1) 
exhibit MCD type. In the O-N plane they are represented with the three asterisks at 
$\log(O/H)+12 \sim 8.1$, $\log(N/H)+12 \sim 8$. The O and N abundances indicate gas 
processed by HBB, typical of massive ($M\sim 6-8~M_{\odot}$) AGB stars; the low oxygen 
suggests a metallicity $Z\sim 4\times 10^{-3}$. The surface helium is not significantly 
enriched, at odds with the expectations regarding massive AGB stars. For what 
regards H1-61 and Th3-4, the observed helium is close to $\sim 11.1$; taking into account 
the errors ($\sim 0.04$), these results can be reconciled with the theoretical expectations 
from SDU computations, thus confirming low-metallicity, massive $\sim 6-7~M_{\odot}$
progenitors for these two PNe. The interpretation for M3-44 is more tricky, because
the observed helium is $\log(He/H)+12 = 9.92$ with an error of $\sim 0.05$; this value
is definitively too low to be compatible with SDU effects, thus ruling out a massive
progenitor. Given the very small argon abundances, our best interpretation is that M3-44
descends from a metal-poor star ($Z\sim 2\times 10^{-3}$), with initial mass 
$\sim 3~M_{\odot}$, which experienced some HBB, triggering the increase in the 
surface N.

Mac 1-11 (PN~G008.6-02.6), H1-46 (PN~G358.5-04.2), M2-50 (PN~G097.6-02.4), He2-62 
(PN~G295.3-09.3) and H1-1 (PN~G343.4+11.9, in the halo) exhibit evidences of the 
presence of silicates in their spectra, despite being in the region of the O-N plane
where we expect to find carbon stars. They are represented by open triangles within
the shaded region in the left panel of Fig.~\ref{fgorny}. 
Mac 1-11 is in the lower region of the shaded region in the left panel of 
Fig.~\ref{fgorny}. Given the uncertainties associated with the measurement of nitrogen
$\sim 0.1$ dex and the poor understanding of the nitrogen enrichment during the RGB
ascending of low-mass stars, we suggest that this PNe descends from a low mass
$\sim 1~M_{\odot}$ progenitor, and is indeed oxygen-rich. 

H1-46 is located in the middle of the C-star region in Fig.~\ref{fgorny}. Either the
N is largely overestimated, or there is no way of explaining this PNe within our
modelling.

\begin{table*}
\caption{Chemical abundances and other properties of {\it Sample 2} PNe}                                       
\begin{tabular}{l l r r r r r  }        
\hline\hline   
 PN~G& Name&  Dust&   Morph.&    log(N/H)+12& log(O/H)+12& log(Ar/H)+12\\\ 
 (1)&(2)&(3)&(4)&(5)&(6)&(7)\\
 \hline \\
000.1+04.3$^1$&    H~1-16&7&$\dots$&7.766$^{+0.107}_{-0.154}$& 8.830$^{+0.145}_{-0.211}$&6.468$^{+0.142}_{-0.173}$ \\ 
000.7+03.2$^1$&  He~2-250&4&$\dots$&8.464$^{+0.214}_{-0.144}$& 8.627$^{+0.233}_{-0.274}$&6.628$^{+0.204}_{-0.186}$ \\ 
001.4+05.3$^1$&    H~1-15&4&$\dots$&7.348$^{+0.225}_{-0.255}$& 8.422$^{+0.992}_{-0.903}$&6.350$^{+0.616}_{-0.444}$ \\ 
001.6--01.3&      Bl~Q&4&$\dots$&8.446$^{+0.067}_{-0.108}$& 8.265$^{+0.053}_{-0.049}$&6.294$^{+0.061}_{-0.038}$ \\ 
001.7--04.6$^1$&    H~1-56&4&$\dots$&7.764$^{+0.139}_{-0.550}$& 8.763$^{+0.090}_{-0.161}$&6.530$^{+0.059}_{-0.092}$ \\ 
002.0--13.4&   IC~4776&7&$\dots$&7.584$^{+0.087}_{-0.019}$& 8.696$^{+0.040}_{-0.075}$&6.320$^{+0.037}_{-0.049}$ \\ 
002.2--02.7$^1$&    M~2-23&6&$\dots$&7.310$^{+0.160}_{-0.213}$& 8.318$^{+0.063}_{-0.099}$&5.980$^{+0.044}_{-0.066}$ \\ 
002.2--09.4$^1$&    Cn~1-5&7&4&8.442$^{+0.325}_{-0.184}$& 8.807$^{+0.083}_{-0.077}$&6.725$^{+0.108}_{-0.094}$ \\ 
002.9--03.9$^1$&    H~2-39&5&$\dots$&7.905$^{+0.096}_{-0.055}$& 8.412$^{+0.073}_{-0.070}$&6.029$^{+0.059}_{-0.057}$ \\ 
003.1+02.9&      Hb~4&7&3&8.648$^{+0.103}_{-0.076}$& 8.699$^{+0.051}_{-0.061}$&6.554$^{+0.033}_{-0.045}$ \\ 
003.2--04.4$^1$&    KFL~12&5&$\dots$&7.045$^{+0.123}_{-0.088}$& 8.889$^{+0.132}_{-0.132}$&  $\dots$   \\ 
003.6+03.1$^1$&    M~2-14&7&$\dots$&8.389$^{+0.073}_{-0.185}$& 8.561$^{+0.083}_{-0.213}$&6.782$^{+0.073}_{-0.155}$ \\ 
004.1--03.8$^1$&    KFL~11&5&$\dots$&7.740$^{+0.082}_{-0.098}$& 8.418$^{+0.094}_{-0.189}$&6.104$^{+0.079}_{-0.122}$ \\ 
006.0+02.8$^1$&    Th~4-3&7&$\dots$&7.722$^{+0.183}_{-1.208}$& 8.365$^{+0.184}_{-2.029}$&5.403$^{+0.113}_{-0.801}$ \\ 
006.0--03.6$^1$&    M~2-31&7&$\dots$&8.360$^{+0.223}_{-0.204}$& 8.662$^{+0.075}_{-0.066}$&6.396$^{+0.084}_{-0.062}$ \\ 
006.1+08.3&    M~1-20&3&1&7.775$^{+0.075}_{-0.077}$& 8.559$^{+0.060}_{-0.077}$&5.994$^{+0.073}_{-0.063}$ \\ 
006.3+04.4$^1$&    H~2-18&5&$\dots$&7.795$^{+0.094}_{-0.087}$& 8.683$^{+0.118}_{-0.068}$&6.167$^{+0.126}_{-0.066}$ \\ 
006.4+02.0$^1$&    M~1-31&7&5&8.563$^{+0.405}_{-0.299}$& 8.701$^{+0.328}_{-0.205}$&6.686$^{+0.222}_{-0.147}$ \\ 
006.5--03.1$^1$&    H~1-61&7&$\dots$&7.985$^{+0.130}_{-0.189}$& 8.053$^{+0.089}_{-0.182}$&6.481$^{+0.085}_{-0.151}$ \\ 
006.8+04.1$^1$&    M~3-15&7&$\dots$&8.364$^{+0.355}_{-0.182}$& 8.723$^{+0.425}_{-0.298}$&6.540$^{+0.454}_{-0.231}$ \\ 
007.2+01.8&      Hb~6&7&2&8.474$^{+0.058}_{-0.116}$& 8.656$^{+0.054}_{-0.086}$&6.571$^{+0.039}_{-0.069}$ \\ 
008.1--04.7$^1$&    M~2-39&4&$\dots$&6.784$^{+0.098}_{-0.057}$& 7.377$^{+0.084}_{-0.042}$&5.314$^{+0.061}_{-0.050}$ \\ 
008.2--04.8$^1$&    M~2-42&0&$\dots$&8.330$^{+0.132}_{-0.080}$& 8.787$^{+0.113}_{-0.039}$&6.338$^{+0.071}_{-0.019}$ \\ 
008.3--01.1&    M~1-40&7&2&8.515$^{+0.046}_{-0.093}$& 8.498$^{+0.053}_{-0.062}$&6.702$^{+0.035}_{-0.053}$ \\ 
008.6--02.6$^1$&  MaC~1-11&5&$\dots$&7.588$^{+0.118}_{-0.092}$& 8.391$^{+0.075}_{-0.064}$&5.891$^{+0.038}_{-0.056}$ \\ 
009.3+04.1&     Th~4-6&6&$\dots$&8.000$^{+0.203}_{-0.514}$& 8.566$^{+0.027}_{-0.220}$&5.989$^{+0.041}_{-0.125}$ \\ 
010.6+03.2&   Th~4-10&1&$\dots$&8.386$^{+0.415}_{-0.325}$& 8.569$^{+0.295}_{-0.554}$&6.477$^{+0.239}_{-0.249}$ \\ 
011.1+07.0&  Sa~2-237&5&4&8.905$^{+0.070}_{-0.108}$& 8.816$^{+0.037}_{-0.100}$&6.655$^{+0.057}_{-0.103}$ \\ 
012.5--09.8&    M~1-62&2&2&6.824$^{+0.174}_{-0.977}$& 8.423$^{+0.060}_{-0.219}$&5.856$^{+0.069}_{-0.083}$ \\ 
014.3--05.5&   V-V~3-6&1&3&7.752$^{+0.250}_{-0.524}$& 8.427$^{+0.023}_{-0.206}$&5.866$^{+0.002}_{-0.155}$ \\ 
018.6--02.2&    M~3-54&0&$\dots$&7.053$^{+0.083}_{-0.129}$& 8.612$^{+0.060}_{-0.053}$&6.336$^{+0.058}_{-0.040}$ \\ 
019.2--02.2&    M~4-10&6&$\dots$&7.779$^{+0.143}_{-0.274}$& 8.576$^{+0.148}_{-0.206}$&6.104$^{+0.079}_{-0.137}$ \\ 
019.4--05.3&    M~1-61&7&5&8.009$^{+0.060}_{-0.060}$& 8.670$^{+0.064}_{-0.096}$&6.380$^{+0.036}_{-0.059}$ \\ 
019.7+03.2&    M~3-25&7&$\dots$&8.215$^{+0.098}_{-0.187}$& 8.740$^{+0.166}_{-0.268}$&6.725$^{+0.135}_{-0.210}$ \\ 
019.7--04.5&    M~1-60&7&$\dots$&8.777$^{+0.106}_{-0.060}$& 8.775$^{+0.074}_{-0.044}$&6.757$^{+0.037}_{-0.034}$ \\ 
020.9--01.1&    M~1-51&7&4&8.572$^{+0.291}_{-0.172}$& 8.811$^{+0.205}_{-0.138}$&6.812$^{+0.219}_{-0.145}$ \\ 
023.8--01.7&    K~3-11&7&$\dots$&7.923$^{+0.062}_{-0.102}$& 8.334$^{+0.072}_{-0.189}$&5.276$^{+0.109}_{-0.102}$ \\ 
025.3--04.6&     K~4-8&5&5&7.785$^{+0.064}_{-0.037}$& 8.592$^{+0.067}_{-0.093}$&6.104$^{+0.032}_{-0.042}$ \\ 
027.6--09.6&   IC~4846&5&2&7.685$^{+0.080}_{-0.115}$& 8.602$^{+0.043}_{-0.130}$&6.134$^{+0.030}_{-0.076}$ \\ 
032.9--02.8&    K~3-19&2&$\dots$&8.250$^{+0.225}_{-0.516}$& 8.579$^{+0.174}_{-0.217}$&6.090$^{+0.105}_{-0.120}$ \\ 
038.7--03.3&    M~1-69&0&$\dots$&8.428$^{+0.118}_{-0.108}$& 8.747$^{+0.057}_{-0.083}$&6.549$^{+0.041}_{-0.055}$ \\ 
042.9--06.9&  NGC~6807&6&4&7.821$^{+0.202}_{--0.020}$& 8.761$^{+0.157}_{-0.064}$&6.149$^{+0.070}_{-0.030}$ \\ 
052.9+02.7&    K~3-31&2&$\dots$&8.193$^{+0.083}_{-0.065}$& 8.989$^{+0.059}_{-0.068}$&6.436$^{+0.054}_{-0.044}$ \\ 
055.5--00.5&    M~1-71&2&$\dots$&8.083$^{+0.103}_{-0.060}$& 8.772$^{+0.071}_{-0.086}$&6.501$^{+0.040}_{-0.049}$ \\ 
060.5+01.8&  He~2-440&4&$\dots$&8.407$^{+0.132}_{-0.050}$& 9.650$^{+0.224}_{-0.067}$&6.573$^{+0.068}_{-0.051}$ \\ 
068.7+01.9&    K~4-41&4&2&8.004$^{+0.105}_{-0.110}$& 8.866$^{+0.079}_{-0.055}$&6.484$^{+0.050}_{-0.046}$ \\ 
079.9+06.4&    K~3-56&5&2&  $\dots$  & 8.228$^{+0.076}_{-0.143}$&5.903$^{+0.040}_{-0.054}$ \\ 
097.6--02.4&    M~2-50&5&2&7.922$^{+0.083}_{-0.069}$& 8.378$^{+0.069}_{-0.063}$&6.083$^{+0.049}_{-0.041}$ \\ 
107.4--02.6&    K~3-87&2&2&  $\dots$  & 8.292$^{+0.099}_{-0.188}$&5.841$^{+0.090}_{-0.067}$ \\ 
184.0--02.1&     M~1-5&2&2&7.447$^{+0.060}_{-0.066}$& 8.083$^{+0.041}_{-0.070}$&5.969$^{+0.041}_{-0.056}$ \\ 
205.8--26.7&   MaC~2$\dots$&2&2&7.328$^{+0.874}_{-0.670}$& 8.320$^{+0.077}_{-0.247}$&5.829$^{+0.061}_{-0.146}$ \\ 
235.3--03.9&    M~1-12&2&5&7.760$^{+0.422}_{-0.578}$& 8.606$^{+0.778}_{-0.966}$&5.501$^{+0.251}_{-0.348}$ \\ 
263.0--05.5&      PB~2&2&2&  $\dots$  & 8.310$^{+0.084}_{-0.074}$&5.905$^{+0.096}_{-0.075}$ \\ 
264.4--12.7$^2$&    He~2-5&2&1&7.554$^{+0.244}_{-0.577}$& 8.378$^{+0.166}_{-0.347}$&5.976$^{+0.168}_{-0.188}$ \\ 
 \hline            
\end{tabular}
\end{table*}

\begin{table*}
\contcaption{Chemical abundances and other properties of {\it Sample 2} PNe}                                       
\begin{tabular}{l l r r r  r r  }        
\hline\hline   
 PN~G& Name&  Dust&   Morph.&    log(N/H)+12& log(O/H)+12& log(Ar/H)+12\\\ 
 (1)&(2)&(3)&(4)&(5)&(6)&(7)\\
 \hline \\

275.3--04.7$^2$&   He~2-21&2&2&7.626$^{+0.035}_{-0.087}$& 8.328$^{+0.076}_{-0.059}$&5.943$^{+0.050}_{-0.041}$ \\ 
278.6--06.7&   He~2-26&2&2&7.997$^{+0.125}_{-0.130}$& 8.519$^{+0.047}_{-0.108}$&6.072$^{+0.038}_{-0.074}$ \\ 
285.4+01.5&    Pe~1-1&1&4&8.049$^{+0.077}_{-0.088}$& 8.607$^{+0.112}_{-0.089}$&6.422$^{+0.058}_{-0.050}$ \\ 
285.4+02.2&    Pe~2-7&0&2&7.438$^{+0.059}_{-0.038}$& 7.484$^{+0.050}_{-0.047}$&5.705$^{+0.050}_{-0.038}$ \\ 
286.0--06.5&   He~2-41&2&4&7.656$^{+0.084}_{-0.053}$& 8.320$^{+0.067}_{-0.074}$&5.957$^{+0.052}_{-0.056}$ \\ 
289.8+07.7$^2$&   He~2-63&0&2&7.724$^{+0.185}_{-0.159}$& 8.425$^{+0.061}_{-0.068}$&5.973$^{+0.145}_{-0.068}$ \\ 
295.3--09.3&   He~2-62&5&4&7.868$^{+0.117}_{-0.067}$& 8.149$^{+0.063}_{-0.066}$&5.923$^{+0.052}_{-0.046}$ \\ 
296.3--03.0&   He~2-73&7&5&8.408$^{+0.082}_{-0.082}$& 8.618$^{+0.063}_{-0.058}$&6.534$^{+0.051}_{-0.051}$ \\ 
297.4+03.7&   He~2-78&1&2&7.871$^{+0.139}_{-0.186}$& 8.677$^{+0.249}_{-0.413}$&6.127$^{+0.127}_{-0.185}$ \\ 
300.7--02.0&   He~2-86&7&4&8.666$^{+0.124}_{-0.146}$& 8.653$^{+0.099}_{-0.152}$&6.729$^{+0.108}_{-0.102}$ \\ 
307.5--04.9&   MyCn~18&7&4&8.281$^{+0.132}_{-0.760}$& 8.590$^{+0.337}_{-1.139}$&6.438$^{+0.187}_{-0.403}$ \\ 
309.0+00.8&   He~2-96&7&2&8.238$^{+0.089}_{-0.108}$& 8.782$^{+0.025}_{-0.114}$&6.666$^{+0.029}_{-0.057}$ \\ 
309.5--02.9&   MaC~1-2&1&2&8.550$^{+0.214}_{-0.257}$& 8.595$^{+0.155}_{-0.255}$&6.420$^{+0.128}_{-0.185}$ \\ 
324.8--01.1&  He~2-133&7&3&8.688$^{+0.192}_{-0.190}$& 8.631$^{+0.172}_{-0.202}$&6.530$^{+0.120}_{-0.135}$ \\ 
327.8--06.1&  He~2-158&7&4&7.967$^{+0.075}_{-0.063}$& 8.449$^{+0.034}_{-0.058}$&6.193$^{+0.041}_{-0.050}$ \\ 
334.8--07.4& SaSt~2-12&4&4&6.844$^{+0.090}_{-0.060}$& 6.656$^{+0.067}_{-0.062}$&  $\dots$   \\ 
336.3--05.6&  He~2-186&1&4&8.364$^{+0.060}_{-0.040}$& 8.588$^{+0.036}_{-0.046}$&6.450$^{+0.030}_{-0.041}$ \\ 
336.9+08.3& StWr~4-10&5&2&7.182$^{+0.099}_{-0.106}$& 8.137$^{+0.054}_{-0.109}$&5.911$^{+0.090}_{-0.106}$ \\ 
340.9--04.6&    Sa~1-5&0&2&7.751$^{+0.063}_{-0.054}$& 8.400$^{+0.057}_{-0.044}$&6.090$^{+0.041}_{-0.031}$ \\ 
343.4+11.9$^2$&     H~1-1&5&3&8.025$^{+0.152}_{-0.157}$& 8.547$^{+0.067}_{-0.073}$&5.820$^{+0.050}_{-0.069}$ \\ 
344.4+02.8&    Vd~1-5&2&$\dots$&7.637$^{+0.065}_{-0.065}$& 8.455$^{+0.057}_{-0.050}$&6.279$^{+0.041}_{-0.037}$ \\ 
345.2--08.8&      Tc~1&3&1&7.498$^{+0.056}_{-0.046}$& 8.453$^{+0.436}_{-0.338}$&6.307$^{+0.395}_{-0.254}$ \\ 
348.4--04.1$^2$&    H~1-21&7&4&8.636$^{+0.135}_{-0.142}$& 8.715$^{+0.078}_{-0.114}$&6.549$^{+0.075}_{-0.109}$ \\ 
348.8--09.0&  He~2-306&0&2&  $\dots$  & 8.718$^{+0.547}_{-0.422}$&6.471$^{+0.467}_{-0.236}$ \\ 
351.9--01.9$^1$& Wray~16-286&7&$\dots$&8.270$^{+0.099}_{-0.109}$& 8.739$^{+0.057}_{-0.074}$&6.442$^{+0.057}_{-0.039}$ \\ 
352.6+03.0$^1$&     H~1-8&7&$\dots$&8.483$^{+0.386}_{-0.201}$& 8.619$^{+0.209}_{-0.141}$&6.635$^{+0.278}_{-0.179}$ \\ 
354.2+04.3$^1$&    M~2-10&1&$\dots$&8.387$^{+0.059}_{-0.088}$& 8.646$^{+0.210}_{-0.199}$&6.542$^{+0.076}_{-0.094}$ \\ 
354.5+03.3$^1$&    Th~3-4&7&$\dots$&8.107$^{+0.109}_{-0.237}$& 8.215$^{+0.108}_{-0.226}$&6.196$^{+0.092}_{-0.173}$ \\ 
355.2--02.5&    H~1-29&7&$\dots$&8.155$^{+0.301}_{-0.208}$& 8.619$^{+0.184}_{-0.092}$&6.320$^{+0.087}_{-0.076}$ \\ 
355.6--02.7$^1$&    H~1-32&6&$\dots$&7.580$^{+0.201}_{-0.137}$& 8.613$^{+0.053}_{-0.101}$&6.220$^{+0.041}_{-0.065}$ \\ 
355.7--03.0$^1$&    H~1-33&1&$\dots$&8.660$^{+0.153}_{-0.170}$& 8.795$^{+0.116}_{-0.166}$&6.524$^{+0.084}_{-0.155}$ \\ 
355.7--03.5$^1$&    H~1-35&6&$\dots$&7.453$^{+0.059}_{-0.248}$& 8.644$^{+0.062}_{-0.123}$&6.286$^{+0.033}_{-0.049}$ \\ 
356.2--04.4$^1$&    Cn~2-2&6&$\dots$&8.303$^{+0.157}_{-0.274}$& 8.750$^{+0.077}_{-0.069}$&6.375$^{+0.074}_{-0.077}$ \\ 
356.9+04.4$^1$&    M~3-38&7&$\dots$&8.500$^{+0.123}_{-0.097}$& 8.378$^{+0.060}_{-0.070}$&6.439$^{+0.043}_{-0.042}$ \\ 
357.6+01.7$^1$&    H~1-23&4&$\dots$&8.512$^{+0.190}_{-0.078}$& 8.798$^{+0.072}_{-0.112}$&6.624$^{+0.066}_{-0.082}$ \\ 
357.6+02.6&    H~1-18&7&$\dots$&9.387$^{+0.402}_{-0.023}$& 9.882$^{+0.884}_{-0.051}$&7.358$^{+0.520}_{-0.013}$ \\ 
358.2+03.6$^1$&    M~3-10&4&$\dots$&8.000$^{+0.049}_{-0.072}$& 8.706$^{+0.065}_{-0.041}$&6.393$^{+0.062}_{-0.042}$ \\ 
358.2+04.2$^1$&     M~3-8&7&$\dots$&8.117$^{+0.145}_{-0.641}$& 8.377$^{+0.364}_{-1.137}$&6.403$^{+0.239}_{-0.676}$ \\ 
358.5--04.2$^1$&    H~1-46&6&$\dots$&7.777$^{+0.067}_{-0.095}$& 8.272$^{+0.065}_{-0.050}$&6.107$^{+0.039}_{-0.042}$ \\ 
358.6+01.8$^1$&     M~4-6&7&2&7.576$^{+0.047}_{-0.052}$& 7.899$^{+0.049}_{-0.075}$&5.994$^{+0.034}_{-0.051}$ \\ 
358.7--05.2$^1$&    H~1-50&7&$\dots$&8.286$^{+0.104}_{-0.063}$& 8.657$^{+0.070}_{-0.053}$&6.371$^{+0.057}_{-0.046}$ \\ 
358.9+03.2$^1$&    H~1-20&7&$\dots$&8.505$^{+0.063}_{-0.078}$& 8.607$^{+0.092}_{-0.101}$&6.600$^{+0.062}_{-0.064}$ \\ 
358.9+03.4$^1$&    H~1-19&7&$\dots$&8.253$^{+0.104}_{-0.164}$& 8.260$^{+0.130}_{-0.291}$&6.553$^{+0.092}_{-0.195}$ \\ 
359.3+03.6$^1$&    Al~2-E&0&$\dots$&8.238$^{+0.111}_{-0.088}$& 8.422$^{+0.068}_{-0.085}$&5.939$^{+0.041}_{-0.055}$ \\ 
359.3--01.8$^1$&    M~3-44&7&$\dots$&8.086$^{+0.074}_{-0.098}$& 8.090$^{+0.146}_{-0.189}$&5.233$^{+0.069}_{-0.087}$ \\ 
359.7--02.6&    H~1-40&7&$\dots$&7.943$^{+0.130}_{-0.325}$& 8.528$^{+0.335}_{-0.401}$&6.233$^{+0.165}_{-0.278}$ \\ 
359.9--04.5$^1$&    M~2-27&7&$\dots$&8.825$^{+0.260}_{-0.117}$& 8.846$^{+0.149}_{-0.106}$&6.746$^{+0.088}_{-0.056}$ \\ 

\hline
\end{tabular}
\end{table*}

The N content of M-50 and He2-62 is too large to be compatible 
with a low-mass progenitor that failed to reach the C-star stage. The discrepancy
between the models and the observation in this case amounts to $\sim 05.$ dex, 
far in excess of the uncertainties associated to the measurements (very small in these
cases) or to the modelling of extra-mixing during the RGB phase. Our favourite
possibility here is that the N is slightly underestimated, thus rendering the
chemical composition of  M-50 and He2-62 compatible with a $\sim 3~M_{\odot}$ progenitor,
that experienced some HBB, thus inhibiting the formation of a carbon star.

For what concerns H1-1, there is the possibility that it descends from a low-mass 
($M<1.5~M_{\odot}$) progenitor, which never reached the C-star stage. The reason for its 
anomalous position, within the C-star region, might be related to a different nitrogen 
content, possibly enhanced, in the halo region where it formed. 

In all these outlier cases, it is almost impossible to proceed without a measure of the 
atomic carbon content of the PNe. With these measurements on hand, we are confident 
that the evolutionary paths to the observed chemistry, given initial mass and metallicity, 
would be more obvious.

\section{Discussion}
Dust properties of PNe have been linked to their progenitors in a variety of environments 
(e.g., Stanghellini et al. 2007, 2012). 

With Sample 1 we find similar conclusions that were addressed by Stanghellini et al. (2007) 
for LMC PNe. 4 out of the 6 PNe for which dust identification is available confirm that
ORD corresponds to $C/O<1$, and CRD to $C/O>1$. Statistics is very limited 
for sample 1 though, and we do find 2 exceptions of CRD PNe with $C/O<1$ (NGC 6826 and NGC 3242).

We can not use Sample 2 PNe for this type of analysis, given the lack of carbon abundances 
for these PNe. 

In this work, using Sample 2 PNe,  we found a similar fraction of subsolar and solar 
metallicity PNe with ORD dust chemistry. GG14 analyzed several of the Sample 2 PNe, 
together with other PNe with known dust properties,  and  found that solar ORD PNe 
predominate. It is worth noting that we did not include in Sample 2 PNe with uncertain 
abundances, while GG14 had, so this selection may explain the different population fraction.

CRD PNe are rare in the bulge, as noted by GG14. On the other hand, MCD PNe are frequent 
in the bulge. These PNe lie in the carbon star progeny locus of the ON diagram (the 
yellow region of Fig.~\ref{fgorny}). Naturally, the conclusion about these progenitors is 
linked to the reliability of the ON diagnostics to predict that these AGB stars do not 
experience HBB. It is also worth adding that, as noted by \citet{garcia16b}, 
we can not disregard other formation channels for MCD PNe, such as extra-mixing, rotation, 
binary interaction, or pre-He enrichment. These channels will only be constrained when 
carbon gas abundances will be available for these PNe.

We found that the majority ($\sim 80 \%$) of disk PNe with sub-solar metallicity are the 
progeny of carbon stars. This results, in agreement with the analysis by GG14, could be 
partly due to the intrinsic difficulty in detecting PNe with very massive progenitors, 
owing to the short time they evolve at high luminosity. This paucity of HBB PNe with 
very massive progenitors is seen in both the disc and the bulge.

By comparing the observed chemistry of each object with the model predictions we found 
that both sub-solar and solar metallicity MCD PNe could be the progeny of massive 
($M >5~M_{\odot}$) AGB evolution. Once again, gas phase carbon abundances will clarify 
this point, especially since there are chemical outliers that could have a very different 
origin (see a relevant discussion in GG14). We also found that the fraction of PNe 
descending from AGB stars that experienced HBB in the bulge is higher ($40\%$) than in 
the disc, suggesting a higher percentage of young PN progenitors in the bulge than in 
the disc. The latter result is based on 12 PNe, thus it must be taken
with some caution, and follow up when carbon abundances of these PNe will be available.

\section{Conclusions}
We study the PNe population of the Milky Way, based on the properties of two samples of 
Galactic PNe, selected according to the availability of their physical parameters in the 
literature. Sample 1 includes 40 PNe, for which the CNO abundances are available, with 
carbon measurements are from ultraviolet emission lines. Sample 2 includes 102 PNe whose 
dust properties and abundances of several elements are available, but atomic carbon has 
not been measured from CELs so far.

By comparing the PNe chemical composition data samples at hand with the yields from AGB 
evolution of an array of models, we discussed the possible progenitors of the PNe observed. 
Particular importance have the abundances of elements related to CNO cycling, which are 
the most sensitive to the efficiency of the physical processes able to alter the surface 
chemical composition of stars during the AGB evolution. 

The mass fraction of neon and argon are also used in the comparison, mainly to infer the 
metallicity of the progenitor populations, because these elements are not expected to 
undergo significant changes during the AGB phase. 

Furthermore, the enrichment in the helium abundance helps the interpretation, being a sign 
of SDU, operating only in massive AGB stars.

According to our interpretation the majority of PN progenitors, about $60\%$, 
have a solar chemical composition, the remaining $40\%$ having metallicities in the range 
$Z_{\odot}/3 < Z < Z_{\odot}/2$. A few metal-poor objects are also present in the samples. 
Half of the sources in both samples disclose a carbon-rich chemistry, thus suggesting a 
C-star origin. These PNe descend from $1.5-3~M_{\odot}$ progenitors, formed between 500 
Myr and 2 Gr ago. 

A small fraction ($20\%$) of  Sample PNe 1 are nitrogen enriched, indicating 
that they have been exposed to HBB during the AGB evolution. The progenitors of these PNe 
are the youngest stars to have ejected a PN, formed not earlier than 250 Myr ago, and with 
mass above $3.5~M_{\odot}$. The fraction of nitrogen rich PNe is slightly higher ($\sim 30\%$)
in Sample 2 than in Sample 1. 

The remaining PNe of both samples are the progeny of low-mass ($M < 1.5~M_{\odot}$) 
stars, which are older than 2 Gyr. These old stars failed to reach the carbon star stage 
because they lost the external envelope before achieving the $C/O>1$ condition at the 
surface of the star.

We conclude that measuring gas phase carbon abundance in PNe is crucial to allow a 
robust classification of PNe and  their progenitors. Carbon is the most sensitive element 
to the two physical mechanisms potentially able to alter the surface chemical composition 
of AGB stars, namely TDU and HBB. In this context, the analysis of Sample 1 PNe provides 
a more robust analysis of the AGB progenitors than that of Sample 2 PNe. The observed 
carbon abundances are nicely reproduced by the yields of AGB evolution used in the 
present analysis, in particular for the upper limits of the carbon amounts. This finding 
supports the outcome of AGB modeling, indicating an upper limit to the quantity of 
carbon which can be accumulated to the surface regions of the stars, and that can be 
observed directly in PNe. This is due to the low effective temperature reached during the 
carbon star stage, which favours the formation of great quantities of carbon dust, leading 
to a rapid loss of the external envelope, once the surface carbon is largely in excess of 
oxygen.

A few Sample 2 PNe present a HBB contaminated chemistry, yet they are unexpectedly are 
surrounded by carbon dust. Measuring atomic carbon from UV CELs in these PNe would be 
extremely important to shed new light on the very final AGB phases of massive AGB stars, 
particularly on the possibility, still highly debated, that late TDU events could favour 
the C-star stage, after HBB is extinguished.

\section*{Acknowledgments}
L.S. is indebted to the Observatory of Rome for the warm hospitality during her sabbatic 
leave and to S.~K. G{\'o}rny for useful discussions. D.A.G.-H. was funded by the 
Ramon y Cajal fellowship number RYC-2013-14182. D.A.G.-H. and F.D. acknowledge support 
provided by the Spanish Ministry of Economy and Competitiveness (MINECO) under grant 
AYA-2014-58082-P.

\end{document}